\documentclass[10pt, conference, letterpaper]{IEEEtran}
\IEEEoverridecommandlockouts
\usepackage[letterpaper, top=0.7 in, bottom=3.25cm, left=0.585 in, right=0.585 in]{geometry}

\usepackage{amsmath}
\usepackage{autobreak}

\usepackage{amssymb,amsfonts}

\usepackage{graphicx}
\usepackage{subfigure} 
\usepackage{epstopdf}
\usepackage{booktabs}
\usepackage{multirow}
\usepackage{algorithm}
\usepackage{algorithmic}
\usepackage{mathrsfs}
\usepackage{bm}
\usepackage{color}
\usepackage{ulem}

\usepackage{cite}
\usepackage{cleveref}

\usepackage{stfloats}
\usepackage{tabularx}
\usepackage{placeins}

\usepackage{amsthm}

\usepackage{cuted}

\Crefname{figure}{Fig.}{Figures}

\allowdisplaybreaks[4]

\begin{document}

\title{Delay-Aware Task Offloading for Heterogeneous VLC-RF-based Vehicular Fog Computing}

\author{
  \IEEEauthorblockN{Nan~An\IEEEauthorrefmark{1}, Hongyi~He\IEEEauthorrefmark{1}, Fang~Yang\IEEEauthorrefmark{1}\IEEEauthorrefmark{2}, Chang~Liu\IEEEauthorrefmark{3}\IEEEauthorrefmark{4}, Jian~Song\IEEEauthorrefmark{5}, Zhu~Han\IEEEauthorrefmark{6}, and Binbin~Zhu\IEEEauthorrefmark{7}}
  \IEEEauthorblockA{\IEEEauthorrefmark{1}Department of Electronic Engineering, BNRist, Tsinghua University, Beijing 100084, China}
  \IEEEauthorblockA{\IEEEauthorrefmark{2}State Key Laboratory of Widegap Semiconductor Optoelectronic Materials and Technologies, Beijing 100084, China}
  \IEEEauthorblockA{\IEEEauthorrefmark{3}The Center of National Railway Intelligent Transportation System Engineering and Technology, Beijing 100081, China}
  \IEEEauthorblockA{\IEEEauthorrefmark{4}Signal \!\&\! Communication \!Research \!Institute, \!China \!Academy \!of \!Railway \!Sciences \!Corporation \!Limited, Beijing \!100081, China}
  \IEEEauthorblockA{\IEEEauthorrefmark{5}Shenzhen International Graduate School, Tsinghua University, Shenzhen 518055, China}
  \IEEEauthorblockA{\IEEEauthorrefmark{6}Department of Electrical and Computer Engineering at the University of Houston, Houston, TX 77004 USA} 
  \IEEEauthorblockA{\IEEEauthorrefmark{7}Shenzhen Hua Chuang Chip Lighting Co., Ltd, Shenzhen 518118, China}

  \vspace{-0.4 in}
}

\maketitle
\begin{abstract}
  Vehicular fog computing (VFC) is a promising paradigm for reducing the computation burden of vehicles, thus supporting delay-sensitive services in next-generation transportation networks. However, traditional VFC schemes rely on radio frequency (RF) communications, which limits their adaptability for dense vehicular environments. In this paper, a heterogeneous visible light communication (VLC)-RF architecture is designed for VFC systems to facilitate efficient task offloading. Specifically, computing tasks are dynamically partitioned and offloaded to idle vehicles via both VLC and RF links, thereby fully exploiting the interference resilience of VLC and the coverage advantage of RF. To minimize the average task processing delay (TPD), an optimization problem of task offloading and computing resource allocation is formulated, and then solved by the developed residual-based majorization-minimization (RBMM) algorithm. Simulation results confirm that the heterogeneous VLC-RF architecture with the proposed algorithm achieves a 15\% average TPD reduction compared to VFC systems relying solely on VLC or RF.
\end{abstract}

\IEEEpeerreviewmaketitle

\section{Introduction}

  The growth of delay-sensitive applications in transportation networks, such as autonomous driving, has intensified the demand for low-delay processing of computation-intensive tasks, which often exceeds the ability of onboard computing resources. Although cloud computing provides abundant resources, its inherent delay renders it unsuitable for delay-critical applications~\cite{cao_mobility-aware_2023}. Besides, while edge computing is geographically close~\cite{nan_joint_2023}, it is hindered by bandwidth limitations and resource competition when serving multiple vehicles simultaneously~\cite{wei_many-many_2024}. To bridge this gap, vehicular fog computing (VFC) has emerged as a promising paradigm for next-generation networks~\cite{wei_ocvc_2023}, where vehicular computing tasks are offloaded to nearby vehicles with idle computing resources, effectively reducing the task processing delay (TPD). Since TPD encompasses computing and transmission delays, both task offloading and transmission have become key issues in VFC systems.
  
  For task offloading, bipartite task partitioning divides computing tasks into two components, which are then processed by onboard central processing unit (CPU), edge servers, or dedicated service vehicles~\cite{feng_joint_2023}. Nevertheless, such a scheme only leveraged the computing resources from a limited number of vehicles, failing to exploit substantial idle resources in the network. In the case of multi-partition offloading, tasks were pre-split and offloaded to multiple idle vehicles based on fuzzy logic~\cite{bute_efficient_2021} and overlapping coalition game~\cite{wei_ocvc_2023}, yet static task partitioning ignored dynamic variations in achievable data rates and available vehicle computing resources, leaving significant optimization potential. Furthermore, dynamic multi-partitioning of tasks and computing resources trading for task execution were enabled via multi-agent reinforcement learning~\cite{wei_many-many_2024}, but its simplified assumption of interference-free parallel transmissions undermines its applicability in realistic multi-vehicle scenarios.
  
  For task transmission, existing VFC implementations predominantly depend on radio frequency (RF) communications for broad coverage, suffering from spectral congestion and competition-induced failures, especially in dense vehicular networks~\cite{karbalayghareh_channel_2020}. In contrast, via directional line-of-sight (LoS) links between headlamps/taillights and photodiodes, visible light communication (VLC) inherently mitigates inter-vehicle interference, which was utilized for task offloading in edge computing~\cite{yang_joint_2024}. Under this circumstance, the heterogeneous VLC-RF architecture aims to combine the interference resilience of VLC and the extended coverage of RF~\cite{memedi_vehicular_2021}, which is expected to reduce the transmission delay and ensure the timely completion of computing tasks.

  The primary contributions of this paper are summarized as follows. Firstly, to the best of our knowledge, it is the first time to integrate heterogeneous VLC-RF communications into VFC, with a comprehensive system model being constructed, where the dynamic multi-partitioning of tasks is enabled to fully utilize the idle computing resources. Secondly, the average TPD minimization problem is formulated, and the corresponding residual-based majorization-minimization (RBMM) algorithm is designed. Thirdly, numerical results demonstrate a 15\% average TPD reduction of the heterogeneous VFC system with the proposed algorithm compared to VFC systems relying solely on VLC or RF.

  In this paper, the system model of heterogeneous VFC is described in \Cref{System Model}. The formulation of the average TPD minimization problem is detailed in \Cref{Average Delay Minimization Problem}, and solved by the RBMM algorithm in \Cref{Task Offloading Optimization}. Numerical simulations are shown in \Cref{Simulation Results}, and finally \Cref{Conclusion} draws the conclusion.

\section{System Model}\label{System Model}

  \begin{figure}[t]
    \centering
    \includegraphics[scale=1.05]{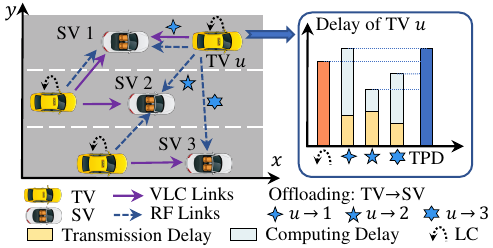}
    \caption{The model of the heterogeneous VLC-RF-based VFC system.}
    \label{fig_scenario}
  \end{figure}

  In \Cref{fig_scenario}, a heterogeneous VLC-RF-based VFC system is considered in transportation network, which comprises $N$ vehicles with width $D_1$ and length $D_2$ driving on a multi-lane road. The spatial coordinate of vehicle $n$ is denoted as $(x_n,y_n)$. While a vehicle is driving, it handles arriving computation-intensive tasks, where the input size and corresponding computation intensity of vehicle $n$ are indicated by $\rho_n$ and $\kappa_n$, respectively. Moreover, the local computing (LC) resource (measured in CPU cycles per second) of vehicle $n$ is indicated by $C_n$.

  To ensure timely task execution, computing tasks are dynamically partitioned between local execution and offloading to multiple idle vehicles via both RF and VLC. In this context, vehicles with task arrivals are designated as task vehicles (TVs) and idle vehicles as service vehicles (SVs), with the set of TVs and SVs represented by $\mathcal{U}=\{1,2,\dots, U\}$ and $\mathcal{V}=\{1,2,\dots, V\}$, respectively. For convenience, a virtual SV $V+1$ is created, where the task offloaded to it is locally computed by the corresponding TV. Consequently, the augmented set of SVs is expressed as $\mathcal{V}_0=\mathcal{V}\cup \{V+1\}$.

  The task partition matrix is defined as $\boldsymbol{M} \in \mathbb{R}^{(V+1) \times U}$, where $\mathbb{R}$ indicates the field of real numbers. Its element $m_{v,u}$ denotes the task proportion offloaded from TV $u$ to SV $v$, and $m_{V+1,u}$ corresponds to the locally processed proportion at TV $u$. Besides, the computing resources allocation matrix is defined as $\boldsymbol{F} \in \mathbb{R}^{V \times U}$, with its element $f_{v,u}$ specifying the proportion of computing resources allocated by SV $v$ to process the task of TV $u$. Therefore, the computing delay for SV $v$ to process the task of TV $u$ is calculated by $Z_{v,u}=m_{v,u} \rho_u \kappa_u/(f_{v,u} C_v)$.

  To reduce TPD, each TV simultaneously offloads tasks to multiple SVs via both RF and VLC, leveraging both communication systems to maintain sustainable offloading during vehicle mobility. To allow concurrent transmission among multiple vehicles, orthogonal frequency-division multiple access is employed in both VLC and RF subsystems, with subchannels denoted as $\mathcal{K}=\{1,2,\dots, K\}$ and $\mathcal{L}=\{1,2,\dots, L\}$, respectively. Furthermore, subchannel assignment is indicated by binary variables $a_{v,u,k}$ and $b_{v,u,l}$. Specifically, $a_{v,u,k}=1$ designates that the subchannel $k$ in the VLC subsystem, while $b_{v,u,l}=1$ describes the subchannel $l$ in the RF subsystem of TV $u$ is allocated to SV $v$, and 0 otherwise.
  
  For RF subsystem, the channel gain on subchannel $l$ between TV $u$ and SV $v$ is denoted by $g_{v,u,l}$ as modeled in~\cite{kai_resource_2019}, and the transmission power of vehicle TV $u$ on subchannel $l$ is signified by $q_{u,l}$, with the collective power matrix as $\boldsymbol{Q}$. Subsequently, the achievable data rate on subchannel $l$ of the RF link between TV $u$ and SV $v$ is represented as 
  \begin{equation}
    \label{RF rate_subchannel}
    S_{v,u,l}= \Gamma_{\mathrm{R}} \log_2\left(1 + \frac{\vert g_{v,u,l} \vert^2 q_{u,l} }{\sum_{i \in \mathcal{U} \backslash u} \vert g_{v,i,l} \vert^2 q_{i,l} + \mu_{\mathrm{R}} \Gamma_{\mathrm{R}}} \right),
  \end{equation}
  where $\Gamma_{\mathrm{R}}$ and $\mu_{\mathrm{R}}$ are the subchannel bandwidth and noise power spectral density (PSD) of the RF subsystem.

  In the VLC subsystem, task offloading is achieved via LoS links between neighboring vehicles, where information is modulated onto the optical intensity, known as intensity modulation and direct detection.  Furthermore, as modeled in~\cite{karbalayghareh_channel_2020,eldeeb_vehicular_2021}, forward and backward VLC channel gains between TV $u$ and SV $v$ are denoted by $h_{1,v,u}$ and $h_{2,v,u}$, respectively. When the LoS link of VLC is blocked, the channel gain is 0. Moreover, the transmission power of vehicle $u$'s headlamps and taillights on subchannel $k$ is denoted as $p_{1,u,k}$ and $p_{2,u,k}$, respectively, with collective power matrices as $\boldsymbol{P}_1$ and $\boldsymbol{P}_2$. Consequently, the achievable data rates on subchannel $k$ of forward and backward VLC links from TV $u$ to SV $v$ are given by
  \begin{equation}
    \label{VLC rate_subchannel}
    R_{\iota,v,u,k} \!=\! \frac{\Gamma_{\mathrm{V}}}{2} \log_2 \! \left( \!1 \!+\! \frac{e}{2\pi} \frac{ h_{\iota,v,u}^2 p_{\iota,u,k} }{\sum_{i \in \mathcal{U} \backslash u} h_{\iota,v,i}^2 p_{\iota,i,k} + \mu_{\mathrm{V}} \Gamma_{\mathrm{V}}} \right) \!,
  \end{equation}
  where $\iota=1$ and 2 indicate forward and backward VLC links, respectively. Besides, $\Gamma_{\mathrm{V}}$ and $\mu_{\mathrm{V}}$ are the subchannel bandwidth and noise PSD of the VLC subsystem.

  Similar to the previous studies~\cite{nan_joint_2023}, the quasi-static method is utilized, in which dynamically arriving tasks are processed in batches over time slots of duration $\Lambda$. In each time slot, the roadside unit collects information from all vehicles, and makes task offloading decisions~\cite{nan_joint_2023,wei_many-many_2024}. Besides, owing to the low relative velocity between co-directional vehicles, their relative positions exhibit minimal variation during the task transmission within each time slot, allowing the VLC channel gain to be treated as constant throughout task transmission~\cite{Chen_time_2016}.

\section{Average Delay Minimization Problem}\label{Average Delay Minimization Problem}

  In the heterogeneous VFC system, the overall achievable data rate of the RF subsystem between TV $u$ and SV $v$ is given by
  \begin{equation}
    \label{RF rate}
    S_{v,u} =  \sum_{l \in \mathcal{L}}  b_{v,u,l} S_{v,u,l}.
  \end{equation}
  Besides, due to the directional propagation of the optical signal, forward and backward VLC links from TV $u$ to SV $v$ cannot coexist simultaneously. Consequently, at most one of the achievable data rates $R_{1,v,u,k}$ and $R_{2,v,u,k}$ on the specific subchannel $k$ is non-zero. Hence, the achievable data rate of the VLC subsystem between TV $u$ and SV $v$ is expressed as
  \begin{equation}
    \label{VLC rate}
    R_{v,u} =  \sum_{k \in \mathcal{K}} a_{v,u,k}  \sum_{\iota=1}^{2} R_{\iota,v,u,k}.
  \end{equation}

  As a consequence, the task transmission delay from TV $u$ to SV $v$ is calculated as
  \begin{equation}
    T_{v,u}=\frac{m_{v,u} \rho_u }{R_{v,u}+S_{v,u}}.
  \end{equation}
  For the sake of convenience, $T_{V+1,u}=0$ is defined for the LC of TV $u$. Besides, the delays of backhaul transmission are ignored due to the relatively small size of output data compared to that of offloaded tasks~\cite{ning_partial_2022}. Therefore, the offloading delay from TV $u$ to SV $v$ is the sum of transmission delay and computing delay, denoted by $T_{v,u}+Z_{v,u}$.

  Since LC and task offloading occur in parallel, the total TPD of TV $u$ is determined by the maximum among its LC delay and its offloading delays of all partitioned tasks. Accordingly, the average TPD among all TVs is given by
  \begin{equation}
    \bar{T}=\frac{1}{U} \sum_{u \in \mathcal{U}} \max_{v \in \mathcal{V}_0} \{T_{v,u}+Z_{v,u}\}.
  \end{equation}

  Consequently, the average TPD minimization problem is formulated as
  \begin{align}
    (\text{P0}): \ \mathop{\min}\limits_{\boldsymbol{M},\boldsymbol{F}}  & \;\; \bar{T} \label{P0}\\ 
    \text{s.t.}   
                  & \;\; m_{v,u} \in [0,1],\ \ \forall \ v \in \mathcal{V}_0, \ \  \forall \ u \in \mathcal{U}, \tag{\ref{P0}{a}}  \label{P0-a}\\
                  & \;\; \sum_{v \in \mathcal{V}_0} m_{v,u} = 1, \ \  \forall \ u \in \mathcal{U}, \tag{\ref{P0}{b}}  \label{P0-b} \\
                  & \;\; f_{v,u} \in [0,1],\ \ \forall \ v \in \mathcal{V}, \ \  \forall \ u \in \mathcal{U}, \tag{\ref{P0}{c}}  \label{P0-c}\\
                  & \;\; \sum_{u \in \mathcal{U}} f_{v,u} \leq 1,\ \ \forall \ v \in \mathcal{V}, \tag{\ref{P0}{d}}  \label{P0-d}\\
                  & \;\; \max_{v \in \mathcal{V}_0} \{T_{v,u}+Z_{v,u}\} \leq \Lambda, \ \  \forall \ u \in \mathcal{U}. \tag{\ref{P0}{e}}  \label{P0-e} 
  \end{align}
  Furthermore, the constraints are detailed as follows.

  \textbf{Task offloading constraints}: Constraints in~\eqref{P0-a} and~\eqref{P0-b} ensure that the entire task of each TV is fractionally partitioned between offloading to SVs and LC.
  
  \textbf{Resource allocation constraints}: The range of computing resource allocation proportion is maintained by constraint in~\eqref{P0-c}. Besides, constraint in~\eqref{P0-d} ensures the allocated resources of an SV are no more than its maximum available amount.

  \textbf{Time slot constraint}: Constraint in~\eqref{P0-e} guarantees the maximum TPD among TVs do not exceed the duration of one time slot.

  In order to address the min-max form in problem (P0), auxiliary variables $\boldsymbol{\zeta}=[\zeta_1, \zeta_2,\dots,\zeta_U]$ are introduced, transforming the original problem into
  \begin{align}
    (\text{P1}): \ \mathop{\min}\limits_{\boldsymbol{M},\boldsymbol{F},\boldsymbol{\zeta}}  & \;\; \frac{1}{U} \sum_{u \in \mathcal{U}} \zeta_u \label{P1}\\ 
    \text{s.t.}   
                  & \;\; \eqref{P0-a}\text{-}\eqref{P0-d}, \nonumber \\
                  & \;\; \zeta_u \leq \Lambda, \ \  \forall \ u \in \mathcal{U}, \tag{\ref{P1}{a}}  \label{P1-a} \\
                  & \;\; T_{v,u} \!+\! Z_{v,u} \leq \zeta_u , \ \  \forall \ v \!\in\! \mathcal{V}_0, \ \  \forall \ u \!\in\! \mathcal{U}. \tag{\ref{P1}{b}}  \label{P1-b} 
  \end{align}
  Consequently, the solution of the problem (P1) satisfies $\zeta_u=\max_{v \in \mathcal{V}_0} \{T_{v,u}+Z_{v,u}\}$, otherwise the $\max_{v \in \mathcal{V}_0} \{T_{v,u}+Z_{v,u}\}$ is a better solution for $\zeta_u$, thus leading to a contradiction. Hence, the equivalence between problems (P0) and (P1) is guaranteed.

\section{Task Offloading and Resource Allocation}\label{Task Offloading Optimization}

  In this section, the constraint in problem (P1) is transformed to a second-order cone (SOC) constraint, and then the reformulated problem is solved by the proposed RBMM algorithm.

  \subsection{Problem Transformation}
  The inherent non-convexity of the problem (P1) poses significant challenges. Accordingly, the fractional constraint in~\eqref{P1-b} is first addressed through the introduction of additional auxiliary variables $\boldsymbol{\Omega_1} \in \mathbb{R}^{V \times U}$ and $\boldsymbol{\Omega_2}\in \mathbb{R}^{V \times U}$, with their elements $\omega_{1,v,u}$ and $\omega_{2,v,u}$, respectively. Accordingly, constraint in~\eqref{P1-b} is decomposed into
  \begin{align}
    & m_{v,u} \rho_u  \leq \omega_{1,v,u}(R_{v,u}+S_{v,u}) , \ \ \forall \ v \in \mathcal{V}, \ \  \forall \ u \in \mathcal{U},  \label{trans-a} \\
    & m_{v,u} \rho_u \kappa_u \leq {\omega_{2,v,u} f_{v,u} C_v} , \ \  \forall \ v \in \mathcal{V},\ \  \forall \ u \in \mathcal{U},  \label{trans-b} \\
    & \omega_{1,v,u} + \omega_{2,v,u} \leq \zeta_u , \ \  \forall \ v \in \mathcal{V}, \ \  \forall \ u \in \mathcal{U},  \label{trans-c} \\
    & m_{V+1,u} \rho_u \kappa_u \leq \zeta_u C_u , \ \  \forall \ u \in \mathcal{U}. \label{trans-d} 
  \end{align}

  To solve the non-convexity of the constraint in~\eqref{trans-b}, a variable substitution is implemented as
  \begin{equation}
    \label{transformation}
    \delta_{v,u}=\sqrt{m_{v,u}} , \ \  \forall \ v \in \mathcal{V}_0,\ \  \forall \ u \in \mathcal{U},
  \end{equation}
  with aggregated substitution matrix denoted as $\boldsymbol{\Delta} \in \mathbb{R}^{(V+1) \times U}$. In this case, constraint in~\eqref{trans-b} is rewritten as
  \begin{equation}
    \delta_{v,u}^2 \rho_u \kappa_u \leq {\omega_{2,v,u} f_{v,u} C_v} , \ \  \forall \ v \in \mathcal{V},\ \  \forall \ u \in \mathcal{U}.
  \end{equation}
  Through further algebraic conduction, it is converted into an equivalent form for $\forall \ v \in \mathcal{V},\ \  \forall \ u \in \mathcal{U}$ as
  \begin{equation}
    \label{SOCP}
    \Vert [C_v f_{v,u} - \omega_{2,v,u}, \ \sqrt{4\rho_u \kappa_u} \delta_{v,u} ] \Vert_2  \leq C_v f_{v,u} + \omega_{2,v,u},
  \end{equation}
  which is a convex SOC constraint. In addition, according to~\eqref{transformation}, the constraints in~\eqref{P0-a},~\eqref{P0-b},~\eqref{trans-a}, and~\eqref{trans-d} are converted to
  \begin{align}
      & \delta_{v,u} \in [0,1],\ \ \forall \ v \in \mathcal{V}_0, \ \  \forall \ u \in \mathcal{U}, \label{P2-a}\\
      & \sum_{v \in \mathcal{V}_0} \delta_{v,u}^2 = 1, \ \  \forall \ u \in \mathcal{U}, \label{P2-b} \\
      & \delta_{v,u}^2 \rho_u  \leq  \omega_{1,v,u}(R_{v,u} + S_{v,u}) , \ \  \forall \ v \in \mathcal{V},\ \  \forall \ u \in \mathcal{U},   \label{P2-c} \\
      & \delta_{V+1,u}^2 \rho_u \kappa_u \leq \zeta_u C_u , \ \  \forall \ u \in \mathcal{U}, \label{P2-d}
  \end{align}
  respectively. Thereafter, problem (P1) is reformulated into
  \begin{align}
    (\text{P2}) & : \ \mathop{\min}\limits_{\boldsymbol{\Delta},\boldsymbol{F},\boldsymbol{\zeta},\boldsymbol{\Omega_1},\boldsymbol{\Omega_2}}   \;\; \frac{1}{U} \sum_{u \in \mathcal{U}} \zeta_u \label{P2}\\ 
    \text{s.t.}   
                  & \;\; \eqref{P0-c}, \ \eqref{P0-d},  \ \eqref{P1-a},  \ \eqref{trans-c},  \ \eqref{SOCP}, \ \eqref{P2-a} , \ \eqref{P2-b}, \ \eqref{P2-c}, \ \eqref{P2-d}. \nonumber
  \end{align}

  \subsection{RBMM Algorithm for Average TPD Minimization}
  In problem (P2), the non-convex equality constraint in~\eqref{P2-b} is addressed through equivalent dual inequality constraints represented as
  \begin{align}
    \sum_{v \in \mathcal{V}_0} \delta_{v,u}^2 &\leq 1, \ \  \forall \ u \in \mathcal{U}, \label{norm-cons1} \\
    \sum_{v \in \mathcal{V}_0} \delta_{v,u}^2 &\geq 1, \ \  \forall \ u \in \mathcal{U}. \label{norm-cons2}
  \end{align}
  While the constraint in~\eqref{norm-cons1} is convex, the non-convex nature of~\eqref{norm-cons2} persists, posing a significant optimization challenge. For the purpose of dealing with this obstacle, a surrogate function is utilized to approximate $\delta_{v,u}^2$ locally, which approximates the feasible region of~\eqref{norm-cons2} by a convex set from inside. Accordingly, to find an appropriate surrogate function of~\eqref{norm-cons2}, the first-order Taylor expansion is applied as $\delta^2 \geq 2\delta_{0}\delta-\delta_{0}^2 $, where the equality is achieved if $\delta=\delta_{0}$. With this foundation, a lower bound of $\delta_{v,u}^2$ is obtained, and then the approximated form of problem (P2) in the $\beta$-th iteration is constructed as
  \begin{align}
    (\text{P3-}\beta) & : \ \mathop{\min}\limits_{\boldsymbol{\Delta},\boldsymbol{F},\boldsymbol{\zeta},\boldsymbol{\Omega_1},\boldsymbol{\Omega_2}} \;\; \frac{1}{U} \sum_{u \in \mathcal{U}} \zeta_u \label{P3}\\ 
    \text{s.t.} 
                  & \;\; \eqref{P0-c}, \ \eqref{P0-d}, \ \eqref{P1-a}, \ \eqref{trans-c}, \ \eqref{SOCP}, \ \eqref{P2-a}, \ \eqref{P2-c}, \ \eqref{P2-d}, \ \eqref{norm-cons1}, \nonumber \\
                  & \;\; \sum_{v \in \mathcal{V}_0} [ 2\delta_{v,u}^{(\beta-1)}\delta_{v,u}-(\delta_{v,u}^{(\beta-1)})^2] \geq 1, \ \  \forall \ u \in \mathcal{U}, \tag{\ref{P3}{a}}  \label{P3-a}
  \end{align}
  where $\delta_{v,u}^{(\beta-1)}$ is the optimization result of the $(\beta-1)$-th iteration.

  To achieve a superior converged solution, the residual approach~\cite{wu_intelligent_2019} is further utilized by introducing residual variables $\boldsymbol{j}_1=[j_{1,1},j_{1,2},\dots,j_{1,U}]$ and $\boldsymbol{j}_2=[j_{2,1},j_{2,2},\dots,j_{2,U}]$, thus transforming problem (P3-$\beta$) into
  \begin{align}
    (\text{P4-}&\beta)  : \ \mathop{\min}\limits_{\boldsymbol{\Delta},\boldsymbol{F},\boldsymbol{\zeta},\boldsymbol{\Omega_1},\boldsymbol{\Omega_2},\boldsymbol{j}_1,\boldsymbol{j}_2}   \;\; \frac{1}{U} \sum_{u \in \mathcal{U}}  \zeta_u + \frac{\Xi}{U} \sum_{u \in \mathcal{U}} (j_{1,u}+j_{2,u}) \label{P4}\\ 
    \text{s.t.}   
                  & \;\; \eqref{P0-c}, \ \eqref{P0-d}, \ \eqref{P1-a}, \ \eqref{trans-c}, \ \eqref{SOCP}, \ \eqref{P2-a}, \ \eqref{P2-c}, \ \eqref{P2-d}, \nonumber \\
                  & \;\; \sum_{v \in \mathcal{V}_0} \delta_{v,u}^2-1 \leq j_{1,u}, \ \  \forall \ u \in \mathcal{U}, \tag{\ref{P4}{a}}  \label{P4-a}\\
                  & \;\; \sum_{v \!\in\! \mathcal{V}_0} \! [ 2\delta_{v,u}^{(\beta-1)}\delta_{v,u} \!-\!(\delta_{v,u}^{(\beta-1)})^2] \!-\! 1  \!\geq\! j_{2,u}, \ \  \forall \ u \!\in\! \mathcal{U}, \tag{\ref{P4}{b}}  \label{P4-b} \\
                  & \;\; j_{1,u} \geq 0, \ \  \forall \ u \in \mathcal{U}, \tag{\ref{P4}{c}}  \label{P4-c}\\
                  & \;\; j_{2,u} \geq 0, \ \  \forall \ u \in \mathcal{U}, \tag{\ref{P4}{d}}  \label{P4-d}
  \end{align}
  where $\Xi$ is the penalty factor. Besides, constraints in~\eqref{P4-a} and~\eqref{P4-b} are converted from~\eqref{norm-cons1} and~\eqref{P3-a} by introducing residuals. Furthermore, because both the objective function and the constraints are convex, the problem (P4-$\beta$) is a convex optimization problem, which can be addressed by the interior point method (IPM)~\cite{boyd2004convex}.
  
  The proposed RBMM algorithm is illustrated in \textbf{\Cref{algorithm-total}}, where the problem (P4-$\beta$) is constructed and solved iteratively until the gap between the average TPDs of two consecutive iterations falls below the threshold. By developing and resolving a sequence of approximated problems iteratively, the optimization result approaches the Karush-Kuhn-Tucker (KKT) point of the original problem~\cite{sun_majorization-minimization_2017}, which underpins the majorization-minimization approach. Furthermore, the computational complexity of IPM in each iteration is $\mathcal{O}((4VU+4U)^{3.5} \log_2(1/\hat{\varepsilon}))$, where $\hat{\varepsilon}$ is the solution accuracy.

  \addtolength{\topmargin}{0.02in}

  \begin{algorithm}[t]
    \caption{RBMM Algorithm for Average TPD Minimization}
    \label{algorithm-total}
    \begin{algorithmic}[1]
      \STATE \textbf{Input}: $\{\rho_u, \kappa_u, C_u, C_v, a_{v,u,k},b_{v,u,l},h_{\iota,v,u},g_{v,u,l} \mid \ \  \forall \ v \in \mathcal{V},\ \  \forall \ u \in \mathcal{U},\ \  \forall \ k \in \mathcal{K},\ \  \forall \ l \in \mathcal{L}, \ \ \forall \ \iota \in \{1,2\} \}$, $\boldsymbol{P}_1$, $\boldsymbol{P}_2$, $\boldsymbol{Q}$, $\sigma_{\mathrm{R}}^2$, $\sigma_{\mathrm{V}}^2$, $\Gamma_{\mathrm{R}}$, $\Gamma_{\mathrm{V}}$, $\Lambda$, and error threshold $\varepsilon>0$.
      \STATE \textbf{Initialization}: $\beta=0$, $\boldsymbol{M}^{(0)}$ and $\boldsymbol{F}^{(0)}$ are initialized, and $\bar{T}^{(0)}$ is calculated according to~\eqref{P1};
      \REPEAT
        \STATE $\beta=\beta+1$;
        \STATE Construct problem (P4) based on $\boldsymbol{M}^{(\beta-1)}$;
        \STATE Solve problem (P4) by IPM and obtain $\boldsymbol{M}^{(\beta)}$ and $\boldsymbol{F}^{(\beta)}$;
        \STATE Calculate $\bar{T}^{(\beta)}$ according to~\eqref{P1};
      \UNTIL{$\vert \bar{T}^{(\beta-1)}-\bar{T}^{(\beta)} \vert < \varepsilon$}
      \STATE $\boldsymbol{M}^{*}=\boldsymbol{M}^{(\beta)}$, $\boldsymbol{F}^{*}=\boldsymbol{F}^{(\beta)}$;
      \STATE \textbf{Output}: $\boldsymbol{M}^{*}$ and $\boldsymbol{F}^{*}$.
    \end{algorithmic}
  \end{algorithm}

  \begin{table}[!tbp]
    \centering
    \renewcommand{\arraystretch}{1.05}
    \caption{Simulation Parameters}
    \label{system_parameter}
      \begin{tabular}{c|c|c}
      \hline\hline
      {\textbf{Type}}           & {\textbf{Parameter}}  & \textbf{Value}    \\ \hline
      \multirow{6}{*}{Vehicle}  & {Vehicle width, $D_1$}                    & 2.2 m             \\ \cline{2-3}
                                & {Vehicle length, $D_2$}                   & 4 m               \\ \cline{2-3}
                                & {Vehicle number, $N$}                     & 40                \\ \cline{2-3}
                                & {Task arrival rate, $\lambda$}            & 10                \\ \cline{2-3}
                                & {Computing resource of a vehicle, $C_n$}  & 2 GHz             \\ \cline{2-3}
                                & {Computation intensity, $\kappa_n$}       & 200 cycles/bit    \\ \hline
      \multirow{2}{*}{System}   & {Time slot duration, $\Lambda$}           & 60 ms             \\ \cline{2-3}
                                & {Penalty factor, $\Xi$}                   & 0.4               \\ \hline
      \multirow{4}{*}{VLC}      & {Subchannel number, $K$}                  & 8                 \\ \cline{2-3}
                                & {Subchannel bandwidth, $\Gamma_{\mathrm{V}}$} & 0.5 MHz       \\ \cline{2-3}
                                & {Transmission power, $p_{\iota,u,k}$}     & 0.1 W             \\ \cline{2-3}
                                & {Noise PSD, $\mu_{\mathrm{V}}$}  & $10^{-21}$ $\text{A}^2$/Hz \\ \hline
      \multirow{4}{*}{RF}       & {Subchannel number, $L$}                  & 8                 \\ \cline{2-3}
                                & {Subchannel bandwidth, $\Gamma_{\mathrm{R}}$} & 0.5 MHz       \\ \cline{2-3}
                                & {Transmission power, $q_{u,l}$}           & 0.1 W             \\ \cline{2-3}
                                & {Noise PSD, $\mu_{\mathrm{R}}$}           & -174 dBm/Hz       \\ \hline
      \end{tabular}
  \end{table}

\section{Simulation Results}\label{Simulation Results}
  In this section, extensive simulations are conducted to evaluate the proposed RBMM algorithm for the heterogeneous VFC system. In the simulation setup, vehicle distribution is modeled as a Poisson distribution, with mobility governed by the intelligent driver model~\cite{feng2021intelligent} across a three-lane road. Besides, the task arrival for each vehicle follows a Poisson process characterized by task arrival rate $\lambda$, while the task size of each TV follows a uniform distribution of $\mathbb{U}[300,500]$ kB. Moreover, each VLC and RF subchannel of a TV is assigned according to the strongest channel gain principle, with transmission power distributed equally across all subchannels. Based on the configuration in~\cite{nan_joint_2023}, the simulation parameters are detailed in \Cref{system_parameter}, which remain unchanged unless otherwise specified.

  \begin{figure}[t]
    \centering
    \includegraphics[scale=0.64]{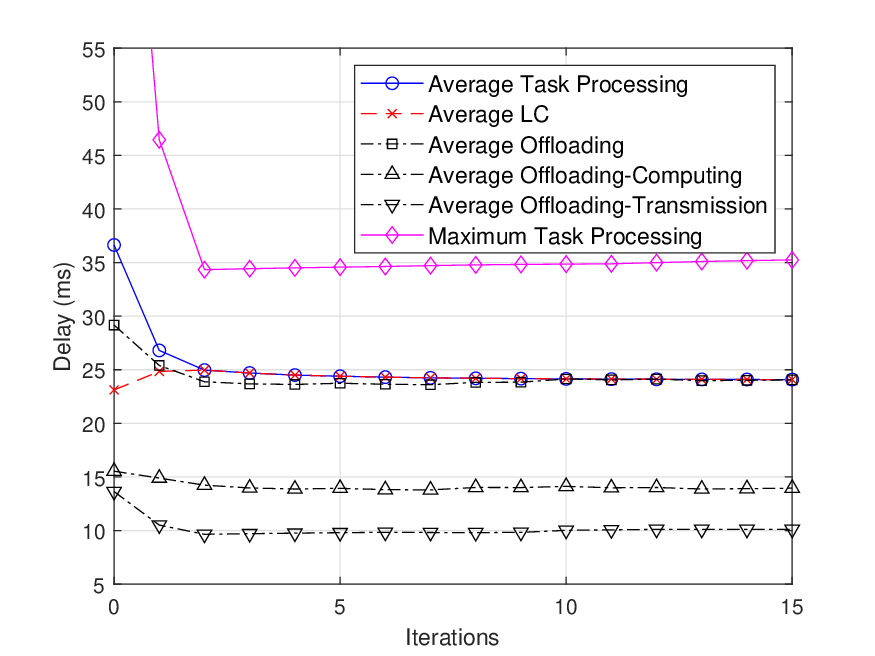}
    \vspace{-0.25in}
    \caption{Various components of TPD during RBMM algorithm convergence.}
    \label{fig_convergence}
  \end{figure}

  \begin{figure}[t]
    \centering
    \includegraphics[scale=0.64]{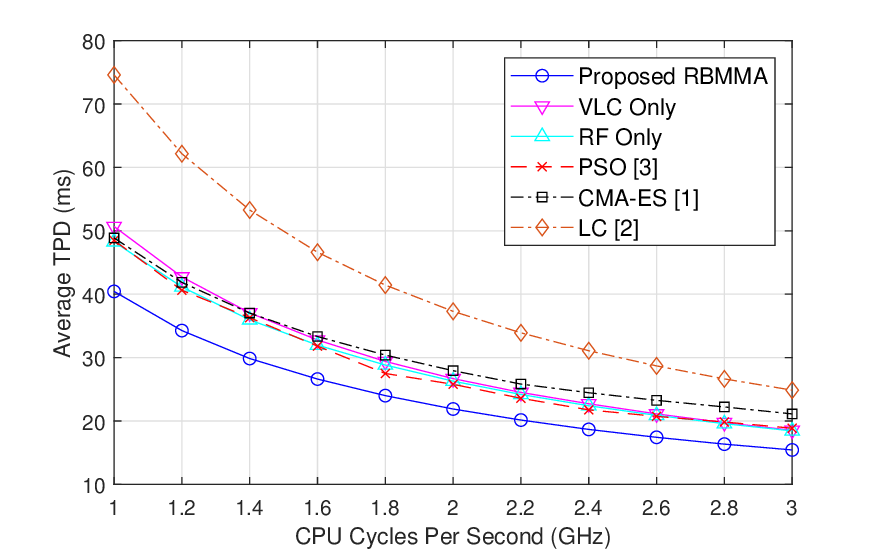}
    \vspace{-0.25in}
    \caption{Comparison of multiple baseline methods under various computing resources.}
    \label{fig_comp}
  \end{figure}

  The convergence characteristics of the proposed RBMM algorithm are illustrated in \Cref{fig_convergence}, where progressive reductions in the average TPD are observed through successive iterations, and the average TPD basically converges at 5 iterations. Besides, compared with the initial iteration point, the proposed RBMM algorithm achieves over 30\% reduction in the average TPD, with even more reductions observed in maximum TPD. Notably, both average LC and offloading delays asymptotically converge toward equilibrium, which arises because the TPD of a TV is governed by the maximum value among its LC delay and parallel offloading delays to multiple SVs. On this occasion, the proposed RBMM algorithm effectively reduces TPD by dynamically balancing these delay components, thereby driving their coordinated convergence.

  \begin{figure}[t]
    \subfigure[Average TPD.]{
      \hspace{-0.2in}
      \includegraphics[scale=0.64]{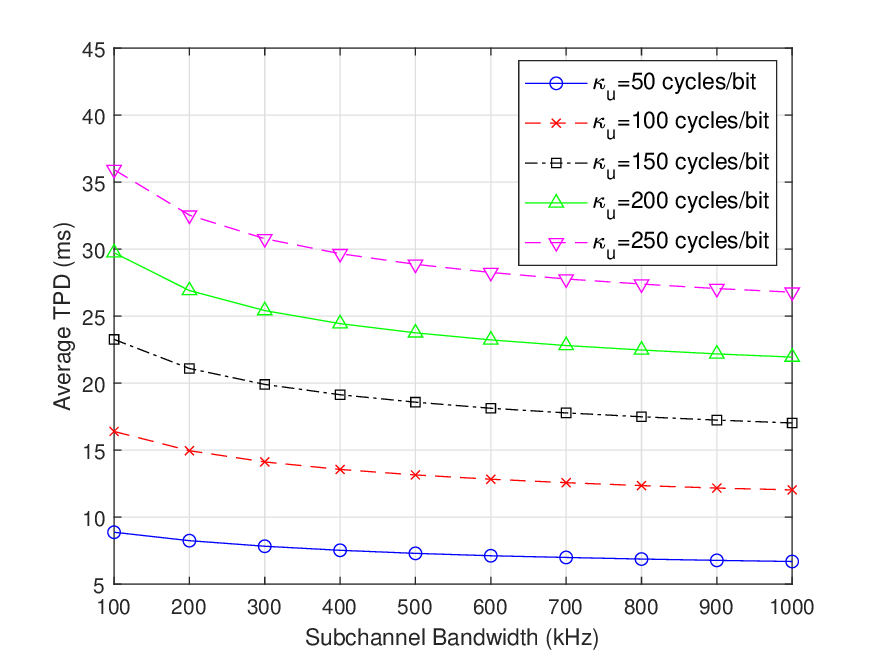}
      \label{fig_comp_bandwidth}}
    \subfigure[Task partition proportion with $\Gamma_\mathrm{R}=\Gamma_\mathrm{V}=500$ kHz.]{
      \hspace{-0.26in}
      \includegraphics[scale=0.64]{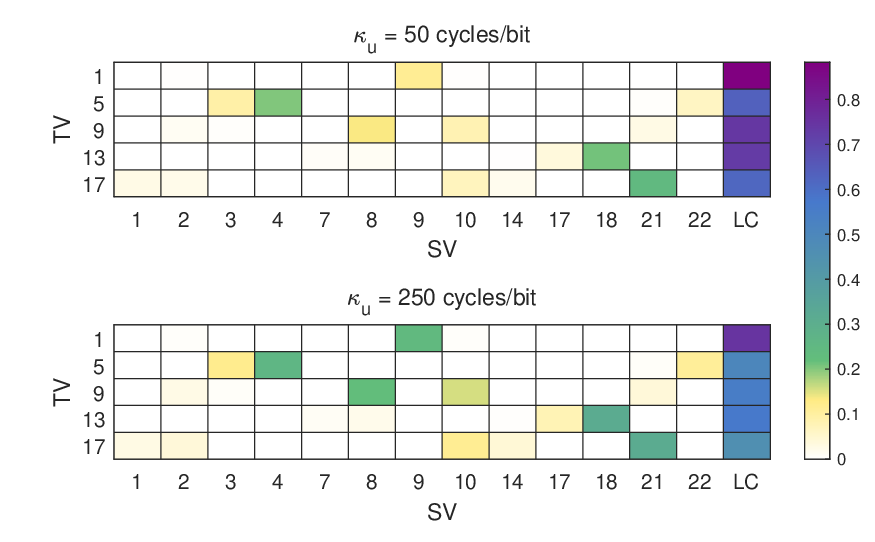}
      \label{fig_task_allo}}
    \vspace{-0.15in}
    \caption{Dual impact of subchannel bandwidth and task computation intensity on the heterogeneous VFC system.}
    \label{fig_comp_band_task}
  \end{figure}

  In \Cref{fig_comp}, the proposed RBMM algorithm is compared with various baselines across varying computing resources. 
  1) \textbf{VLC only}: Tasks are offloaded only via VLC. 
  2) \textbf{RF only}: Tasks are offloaded only via RF.
  3) \textbf{LC~\cite{nan_joint_2023}}: All tasks are computed by the corresponding TV locally without offloading.
  4) \textbf{Particle swarm optimization (PSO)~\cite{wei_many-many_2024} for heterogeneous VFC}: The high-quality optimization solution is found through the cooperation and information sharing in the particle swarm.
  5) \textbf{Covariance matrix adaptation-evolution strategy (CMA-ES)~\cite{cao_mobility-aware_2023} for heterogeneous VFC}: By dynamically adjusting the covariance matrix to guide the search direction and step size of the evolution strategy, a high-quality solution is achieved.

  The proposed heterogeneous VLC-RF architecture achieves an approximately 15\% reduction in the average TPD compared to systems with only VLC or RF, validating its efficacy in VFC. In contrast to LC, the heterogeneous architecture with the proposed RBMM algorithm reduces the average TPD by over 40\%, thereby demonstrating the significant potential of VFC in mitigating TPD. Furthermore, the RBMM algorithm exhibits a 15\% superiority in the average TPD over PSO and CMA-ES methods, highlighting its effectiveness in heterogeneous VFC systems. Additionally, increasing the computing resources of each vehicle is shown to produce monotonic TPD reductions, accompanied by gradual convergence between the average TPD and average LC delay. This convergence is attributed to the significant reduction in computing delay achieved through enhanced computing resources, thereby rendering transmission delay the dominant contributor to offloading delay. Consequently, a larger proportion of tasks is retained for LC by TVs, effectively mitigating the transmission delay, and balancing the local and offloading delays. This equilibrium ultimately lowers the average TPD, resulting in an asymptotic convergence between the average TPD and average LC delay.

  In \Cref{fig_comp_band_task}, the dual impact of task computation intensity and subchannel bandwidth is elucidated. As demonstrated in \Cref{fig_comp_bandwidth}, when subchannel bandwidth increases, the achievable data rates of both VLC and RF are improved, thereby reducing the transmission delay and inducing a consistent decline in the average TPD. Conversely, higher task computation intensities necessitate more resources for computing, leading to prolonged computing delay and consequently elevating the average TPD. Notably, under high task computation intensity conditions, subchannel bandwidth expansion yields more pronounced reductions in the average TPD. As the computation intensity increases, computing delay is amplified, which necessitates a larger proportion of tasks offloading from TVs to SVs to balance LC and offloading delays. In this case, the bandwidth expansion effectively mitigates the transmission delay induced by the increased offloading tasks under high computation intensity, thereby further reducing average TPD. Concurrently, these analytical insights are further validated in~\Cref{fig_task_allo}, where the task partition proportions of some TVs are demonstrated. When the task computation intensity increases from 50 to 250 cycles/bit, a reduction of at least 13\% in the proportion of locally computed tasks per TV is observed, where the reduced part of the tasks is offloaded to other SVs, thus effectively reducing the average TPD. Besides, because offloading delay includes transmission delay in addition to computing delay, the aforementioned balancing mechanism between LC and offloading delays leads to a high proportion of tasks being processed locally.

\section{Conclusion}\label{Conclusion}
  In this paper, a heterogeneous VLC-RF-based VFC system was proposed, enabling the dynamic task multi-partitioning and offloading to idle SVs via both VLC and RF. Based on the constructed comprehensive system model, the average TPD minimization problem was formulated and solved by the proposed RBMM algorithm, which jointly optimized task offloading and computing resource allocation. Furthermore, numerical simulations demonstrated that the heterogeneous VFC system with the proposed RBMM algorithm achieved a 15\% average TPD reduction compared to VLC-only or RF-only VFC systems. Moreover, the proposed algorithm demonstrated a significant reduction in average TPD relative to baseline methods, highlighting its potential to support delay-sensitive services in next-generation transportation networks. In future work, allocation of both communication and computing resource can be optimized with task offloading to enable efficient VFC.

\section*{Acknowledgement}
This work was supported in part by the Shenzhen Municipal Government under Grant SZXJP20230703093002006; in part by the NSF ECCS-2302469; in part by JST ASPIRE JPMJAP2326, and in part by Amazon.

\normalem
\bibliographystyle{IEEEtran}
\bibliography{refs}

\end{document}